\documentclass[aps,showpacs,pra]{revtex4}
\usepackage{epsfig,amssymb,amscd}
\begin{document}
\bibliographystyle{prsty}

\title{\bf Spin-based quantum gating with semiconductor quantum dots by bichromatic radiation method}
\author{Mang Feng$^{1,2}$, Irene D'Amico$^{1}$, Paolo Zanardi$^{1}$ and Fausto Rossi$^{1,3}$}    
\affiliation{$^{1}$ Institute for Scientific Interchange (ISI) Foundation,\\ 
Villa Gualino, Viale Settimio Severo 65, I-10133, Torino, Italy\\
$^{2}$ Laboratory of Magnetic Resonance and Atomic and Molecular Physics,\\ 
Wuhan Institute of Physics and Mathematics, Chinese Academy of Sciences, Wuhan, 430071, China \\
$^{3}$ Dipartimento di Fisica, Politecnico di Torino, Corso Duca degli Abruzzi 24-10129 Torino, Italy} 
\date{\today}

\begin{abstract}

A potential scheme is proposed for realizing a two-qubit quantum gate in semiconductor quantum dots.
Information is encoded in the spin degrees of freedom of one excess conduction electron of each quantum dot. 
We propose to use two lasers,  radiating two neighboring QDs, and tuned to blue detuning with respect
to the resonant frequencies of  individual excitons. The two-qubit phase gate can be achieved by means of 
both Pauli-blocking effect and dipole-dipole coupling between intermediate  excitonic states. 
\end{abstract}
\vskip 0.1cm
\maketitle

Quantum computing with semiconductor quantum dots (QDs) has  drawn more and more interest over the past few 
years \cite{review}. Besides the ease of scalability, semiconductor QD quantum computing is more appealing than 
other quantum computing schemes due to the existence of the industrial basis for semiconductor processing as
well as the promise of being easily integrable. 

In the schemes for quantum computing with semiconductor QDs proposed so far, either the excitonic or  spin
degrees of freedom  have been identified as potential qubits. Quantum gating based on excitons, although being
strongly restricted by the  short decoherence time of the exciton,
can be implemented ultrafastly with optical manipulations \cite{biolatti}.
In contrast, spin qubits, due to their relatively long decoherence 
time \cite {cor, imamo}, allow for  longer storage of quantum information.
Quantum gates can be carried out by means of Coulomb interaction \cite{biolatti,loss,pazy,cal} or by coupling to a cavity mode 
\cite{imamo,sherw,feng1}. When nearest-neighbor coupling plays an  important role 
one has to face a significant overhead  for coupling two distant QDs. On the contrary  if the QDs are
put into a cavity, two distant QDs can interact directly through coupling to the same cavity mode. 
Nevertheless, the implementation time in such a model is typically quite long due to both the 
large detuning technique adopted for avoiding the cavity decay and the weak cavity-laser-QD coupling 
\cite{feng1}. 

More and more experimental evidences, based on current state-of-art nanostructure and laser technology, have shown the present capacity 
to manipulate semiconductor QDs based qubits. 
The single QD cooled and prepared with  one excess conduction band electron 
only, which is the prerequisite of spin-based
QD quantum computing schemes, has already been achieved \cite {fin}. Entangled excitonic states have been realized
by optical method \cite{chen} and the ultrafast spin rotation by laser pulses  in a magnetic field has been presented\cite{gup}. 
More recently, the writing and readout operations for the spin states of the single conduction band electron were performed in an 
n-doped InAs-GaAs QD by nonresonant circularly polarized optical pumping \cite {cor}. In the same experiment, a long lifetime of the 
electron spin was observed.  Although no experiment  so far has demonstrated a quantum algorithm, the experimental 
progress outlined above has paved the road to a working quantum computer with semiconductor QDs.

The present work concentrates on an alternative two-qubit proposal for spin qubits of semiconductor QDs, which is inspired by 
\cite {pazy}. The essential ingredient of our gate proposal is provided  by the  bichromatic radiation approach widely employed 
in ion trap quantum computing \cite{sm}. 
As in \cite {pazy}, quantum information in our scheme will be encoded in the spin states 
of the single excess conduction electron of the QD and the two-qubit quantum gate will be realized by 
exploiting the biexcitonic shift due to the dipole-dipole interaction between adjacent excitonic states. 
However, {\it at variance from} \cite {pazy},  the two-qubit gate  proposed here
is implemented by  the bichromatic radiation approach. 
To this aim, we have to modify the original formulation used for 
trapped ions; this is  due to the differences between QDs and atomic ions, 
i.e., degeneracy of  qubit states in QDs in the absence 
of external field, no metastable levels to use as ancillary states in QDs,
 no motional degrees of freedom  attached to the qubit states and no exactly identical self-assembled QDs. 

Consider two lasers, radiating two neighboring QDs, and tuned to blue detuning with respect
 to the resonant frequencies of  individual excitons. If the sum 
of the two blue detunings equals to the biexcitonic shift, an effective coupling  can  be generated  between 
$|11\rangle_{ab}$ and $|XX\rangle_{ab}$, where $|1\rangle_{k}$ and $|X\rangle_{k}$ denote a qubit state and the 
excitonic state of the QD $k$ respectively (defined later, see Fig. 1). 
We shall show how to realize a conditional phase gate,  based on this effective coupling,
 by means of properly tailored ultrafast laser pulses. 
We shall focus on the case of two QDs. Both our method and results can be  in principle
extended to multi-QD systems. 

Let us suppose that each QD contains only one excess conduction-band electron.  We employ the spin 
states $m_{z}=1/2$ and -1/2 of such electron as qubit states $|1\rangle$ and $|0\rangle$ 
respectively.  Excitonic states are introduced as ancillary ones. Besides the Coulomb repulsion, the 
Pauli-blocking mechanism is essential to our scheme. When we radiate a $\sigma^{-}$ polarized light with suitable energy 
on the QD, due to the Pauli exclusion principle, the exciton $|m_{J}^{e}=-\frac {1}{2}, m_{J}^{h}=-\frac {1}{2}\rangle$ 
in the s-shell \cite{exp1} will be produced if and only if the excess 
electron has a spin projection 1/2. This Pauli-blocking mechanism has been used to experimentally produce 
entangled excitonic states in \cite {chen}. We define 
$|0\rangle_{\nu}=c^{\dagger}_{\nu,-\frac {1}{2}}|vac\rangle$, 
$|1\rangle_{\nu}=c^{\dagger}_{\nu,\frac {1}{2}}|vac\rangle$, and the excitonic states
$|X\rangle_{\nu}=c^{\dagger}_{\nu,-\frac {1}{2}}c^{\dagger}_{\nu,\frac {1}{2}}
d^{\dagger}_{\nu,-\frac {1}{2}}|vac\rangle$, 
where $c^{\dagger}_{\nu,\sigma}(d^{\dagger}_{\nu,\sigma})$
is the creation operator for a conduction (valence) band electron (hole) 
in QD $\nu$ with spin projection $\sigma$, and $|vac\rangle$ denotes the electron-hole vacuum. 
The general Hamiltonian of such a system can be found in \cite{pazy}. Here we only consider a special
situation, i.e., two neighboring QDs with different configurations, radiated by two blue-detuned lasers simultaneously, 
as shown in Fig. 1. 
The Hamiltonian of the QD system $H_{0}+H_{I}$ can be written in unit of $\hbar$ as
\begin{equation}
H_{0}= \Delta | XX \rangle\langle XX | +  \omega_{a} | X \rangle_{a}\langle X |\otimes\hat{I}_{b} + 
\omega_{b} \hat{I}_{a}\otimes| X \rangle_{b}\langle X |
\end{equation}
and
\begin{equation}
H_{I}= \frac {1}{2} [\Omega_{a}(t) \left (e^{i\omega_{L1}t} + e^{i\omega_{L2}t}\right )|1\rangle_{a}\langle X|\otimes\hat{I}_{b} +
\Omega_{b}(t) \left (e^{i\omega_{L1}t}+e^{i\omega_{L2}t}\right )\hat{I}_{a}\otimes |1\rangle_{b}\langle X| + h.c.]
\end{equation}
where $\Omega_{k}(t)$ ($k=a$ and $b$) denotes the couplings of lasers with QDs $a$ and $b$, respectively, 
For simplicity, we assume here that the coupling strength $\Omega_{k}(t)$ is identical for each QD irradiated by different laser beams. 
But this assumption is not essential to the follwoing deduction.
$\hat{I}_{k}$ is the identity operator with respect to QD $k$, $\omega_{k}$ is the resonant energy of a single exciton 
produced in individual QDs, and $\Delta$ is the biexcitonic shift due to Coulomb repulsion. $\omega_{Ln}$ $(n=1$ or $2)$ is 
the laser frequency applied on QDs $a$ and $b$, and $h.c.$ means hermitian conjugate. The Pauli 
blocking is reflected by the absence of the transition from $|0\rangle$ to $|X\rangle$. In the rotating frame with respect to $H_{0}$, 
we have 
$$H' = \frac {\Omega_{a}(t) }{2} \left (e^{i\omega_{L1}t} + e^{i\omega_{L2}t}\right ) e^{-i\omega_{a}t} |1\rangle_{a}\langle X|
\otimes \left (|X\rangle\langle X|e^{-i\Delta t}+|1\rangle\langle 1|+|0\rangle\langle 0| \right )_{b}$$
\begin{equation}
+ \frac {\Omega_{b}(t) }{2} \left (e^{i\omega_{L1}t} + e^{i\omega_{L2}t}\right ) e^{-i\omega_{b}t} 
 \left (|X\rangle\langle X|e^{-i\Delta t}+|1\rangle\langle 1|+|0\rangle\langle 0| \right )_{a}
\otimes|1\rangle_{b}\langle X| + h.c.
\end{equation}
Since the two QDs are radiated by two lasers simultaneously, there should be four detunings. We define 
$\delta_{a}=\omega_{L1}-\omega_{a}$, $\delta^{'}_{a}=\omega_{L2}-\omega_{a}$, $\delta_{b}=\omega_{L1}-\omega_{b}$, and 
$\delta^{'}_{b}=\omega_{L2}-\omega_{b}$ with $\delta_{a}+\delta^{'}_{b}=\delta_{b}+\delta^{'}_{a}=\Delta$ to achieve the 
two-photon resonance. If we adjust the two lasers to satisfy $\Omega_{k}/2 \ll
min\{\delta_{a}, \delta_{b}, \delta^{'}_{a}, \delta^{'}_{b}\}$, then there would be no actual excitation in the intermediate states 
$|1X\rangle_{ab}$ and $|X1\rangle_{ab}$. We thus reach an effective Hamiltonian 
\begin{equation}
H_{eff}=\frac {\tilde{\Omega}(t)}{2} (|XX\rangle\langle 11| + h.c.)_{ab}
\end{equation}
with
$$\frac {\tilde{\Omega}(t)}{2} = \frac {1}{2}\Omega_{a}(t)\Omega_{b}(t)(1/\delta_{a} + 1/\delta^{'}_{a} + 1/\delta_{b} + 
1/\delta^{'}_{b})$$
\begin{equation}
=\frac {1}{2}\Omega_{a}(t)\Omega_{b}(t)\left [ 1/\delta_{a} + 1/(\Delta+\delta-\delta_{a}) + 1/(\delta_{a}-\delta) + 
1/(\Delta-\delta_{a})\right ]
\end{equation}
where $\delta=\omega_{b}-\omega_{a}$ is the resonance frequency difference between the two QDs. 
Based on Eq. (4), returning to the Schr\"odinger representation, we have the time evolution as follows
\begin{equation}
|11\rangle\rightarrow \cos [\frac {1}{2}\int_{0}^{T}\tilde{\Omega}(t)dt] |11\rangle - 
ie^{-i(\omega_{a}+\omega_{b}+\Delta)t}\sin [\frac {1}{2}\int_{0}^{T}\tilde{\Omega}(t)dt] |XX\rangle
\end{equation}
and a similar equation for $|XX\rangle$.

Our conditional phase gate is based on Eq. (6), which can lead to a universal quantum computing along with
single qubit operations. Since our computational subspace is spanned by $|0\rangle_{a(b)}$ and 
$|1\rangle_{a(b)}$, an evolution with $\int_{0}^{T}\tilde{\Omega}(t)dt = 2\pi$ yields 
$|11\rangle_{ab} \rightarrow -|11\rangle_{ab}$, but no change in $|10\rangle_{ab}$, $|01\rangle_{ab}$ and 
$|00\rangle_{ab}$. This is a typical conditional phase gate. 
During the gating, however, the excitonic states are actually excited. As a result, our gating time must 
be shorter than the decoherence time of the exciton. 
Let us suppose that the QDs are made of III-V semiconductor materials. The biexcitonic shift $\Delta$ between
adjacent excitons, corresponding to the inter-dot distance of 10 nm in the presence of an in-plane electric field $F=75 kV/cm$, is about 
$\Delta=4$ $meV$ \cite{biolatti}. We assume $\delta=1$ meV and
the laser pulses to be Gaussian where $\Omega_{k}(t)=\Omega_{k} e^{-t^{2}/2\tau^{2}}$ with $\tau$ the
pulse duration and $\Omega_{a}\approx\Omega_{b} = \Omega_{0}$. Then we have to satisfy the relation
\begin{equation}
\frac {1}{2}\Omega^{2}(0)\left [ 1/\delta_{a} + 1/(5-\delta_{a}) + 1/(\delta_{a}-1) + 
1/(4-\delta_{a})\right ]\int_{-T/2}^{T/2} e^{-t^{2}/\tau^{2}}dt=\pi
\end{equation} 
where the implementation time of the conditional phase gate $T$ should be shorter than the dephasing time 
of the excitons, which is of the order of 1 $ns$ \cite {bayer}. To analyze the constraints given by the short decoherence time
of the excitons and the virtual excitation of the intermediate states, let us define $R=\Omega_{0}/2\delta_{min}$
with $\delta_{min}=min\{\delta_{a},\delta_{b},\delta^{'}_{a},\delta^{'}_{b}\}$. The numerical results in Fig. 2 demonstrate that 
when $\delta_{a}= 2.5$ meV, we have shortest gating times \cite {exp2}. So in what follows, for simplicity, we shall 
only consider this optimal case. If we consider $R=1/2$,
which was adopted in \cite{ion} for building entangled states of trapped ions based on the proposals of \cite{sm},
then $\tau\approx 1.0$ $ps$. But this R is too big to carry out our scheme with high fidelity. To avoid the excitation in the 
intermediate states defined as $\bar{n}=2R^{2}$\cite {ion}
though, we have to restrict $R$ to be smaller than 1/7, i.e., $\bar{n}< 5\%$. Thus to realize a
gating with such a high fidelity, we would have $\tau\approx 9.5$ $ps$.

As mentioned above, our scheme is rooted in \cite{pazy}. Besides using the same qubits, both the
schemes perform the quantum gating by means of the biexcitonic shift and Pauli blocking. 
 The  important difference
is that,  instead of an adiabatic process for accumulating the conditional phase factor  \cite{pazy},
our conditional phase gate is based on the resonant transition between $|11\rangle$ and $|XX\rangle$. This can be achieved
by one-step implementation, which much simplifies the operations in the original proposal of the
bichromatic radiation \cite{feng2}. Nevertheless, due  both to detunings to the 
individual excitation of the exciton and to the second-order process (Eq. (5)) employed in our scheme,
the coupling of the lasers to the QDs cannot be large. As a result, 
the implementation time our gating takes is of the same order of in \cite{pazy}. On the other hand, 
 since the bichromatic radiation approach has been proven experimentally \cite {ion}
in atomic physics to be an efficient and reliable way of entangling states with high fidelity, as 
long as $\Omega_{k}/2 \le \frac {1}{7} min\{\delta_{a}, \delta_{b}, \delta^{'}_{a}, \delta^{'}_{b}\}$ is satisfied, we are optimistic 
on the possibility to achieve a conditional phase gate with the fidelity higher than $95\%$.

Besides the two-qubit gate, single-qubit operation is necessary for a universal quantum computing scheme. As done in 
\cite {biolatti,pazy}, based on the exact knowledge of specific QDs, we assume the individually addressing of the QDs is available 
with laser beams by using energy selective schemes rooted in the characteristic size fluctuations 
of self-assembled QDs combined with near field technique. By employing resonant Raman 
coupling  between $|0\rangle$ and $|1\rangle$ under the radiation of two lasers with different 
polarizations and suitable frequencies \cite{imamo,feng1}, single-qubit rotation can be readily carried out within the order of $ps$.
However we noticed that the light hole $|m_{J}^{h}=1/2\rangle$ is an excited state in III-V semiconductor materials, whose decoherence 
is not advantageous to our single-qubit gating. To avoid this problem, we can choose II-V semiconductor QDs, in which the light hole 
state is energetically favoured \cite{perez}.  

The readout of the final state may be performed again 
via Pauli-blocking schemes: only if the final state is $|1\rangle$,  an exciton may be induced by a 
$\sigma^{-}$ polarized laser pulse of suitable frequency. Therefore, a  $\sigma^{-}$ polarized photon will be created after the exciton
decays. By detecting this photon, we shall know whether the QD spin state is in $|1\rangle$ or $|0\rangle$. 
Since in the readout stage the spins are in product states, we only need to consider the lifetime (i.e., $T_{1}$) of the spin state.
The lifetime of $|1\rangle$ is of the order of $\mu s$. So we can repeat 
this laser pulse excitation 
for thousands of times, which is very similar to the electronical shelving amplification used in ion trap experiments \cite {ion,nag}. 
Although the detection efficiency in our scheme would be somewhat lower than that in a microcavity \cite {imamo,feng1} due to the 
finite angle coverage of the detector, the information amplification mentioned above can guarantee our readout to be correct and 
effective.

However, like previous proposals \cite {biolatti,loss,pazy}, our conditional phase gate is based on a 
nearest-neighbor coupling, which need significant overhead for coupling two distant qubits, and like 
in \cite {biolatti,pazy}, the external electric field is necessary in
our scheme to enlarge the biexcitonic shift $\Delta$. On the contrary, in our other scheme \cite{feng1}, the 
qubits based on the QDs embedded in a high-Q single-mode cavity enjoy an effective coupling between two 
non-neighboring QDs through coupling to the same cavity mode and also no need of external field. 
Nevertheless, in the present scheme, the conditional two-qubit gate can be
carried out more quickly than in \cite{feng1}, which is of great importance in view of decoherence.
The scheme in \cite{feng1} is strongly restricted by the number of the QDs available in a cavity. It is
still experimentally challenging to have few QDs in a high-Q cavity with desired couplings. On the contrary, the
present scheme is more easily scalable.

The quantum gate based on our scheme can be carried out with high fidelity. The decoherence time of the spin state of
the conduction band electron can be of the order of $\mu s$ \cite {imamo}, which is much longer than the decoherence time of the 
exciton and thereby will not affect our gating. If we neglect in our discussion any errors due to incorrect or inappropriate operations, 
potential error sources for our scheme are from (1) actual excitation of the intermediate states and the spontaneous emission 
from excitonic states; (2) small admixture of heavy hole component to the light hole wavefunction due to the interaction
between the hole bands in actual QDs \cite {lut}; (3) possible spectral diffusion due to strong built-in fields and many-body effect; 
(4) the F{\"o}rster process \cite{quiroga} happening in the nearest-neighbor coupled QDs. As discussed above,  error (1) can be 
highly suppressed by reasonable laser-QD coupling and implementation time.  Error (2) yields slight violation of the Pauli blocking,
i.e., a partial excitation of the excitonic state $|m^{e}_{J}=1/2, m^{h}_{J}=-3/2\rangle$ produced in each radiation with the 
$\sigma^{-}$ polarization 
when the spin projection of the only excess conduction electron is $-\frac {1}{2}$. But this can be avoided in our single-qubit gate 
by using in-plane directed laser pulses.  Because of the restriction from symmetry, the induction of
the heavy hole part is prohibited in the mixed wave function for any radiation along the growth direction \cite {weiner}. 
In the implementation of our two-qubit phase gate instead, in order to avoid decoherence related to light-heavy hole mixing,
 it is possible to resort to  adiabatic techniques \cite {pazy, cal}.
Error (3) would result in random level shift and cross biexcitons. Fortunately, in the low temperature as considered in this work,
it happens on the timescale of seconds \cite{robin}, much longer than our gating time. So we can neglect it.
Error (4) will be also greatly suppressed because of the energy spectrum natural mismatch between different QDs.

In conclusion, a bichromatic radiation scheme for implementing two-qubit phase gate with semiconductor QDs 
has been proposed. Our scheme can be considered as a combination of the 
Pauli-blocking spin-based quantum gating and the bichromatic radiation approach.
In principle, both the method and the results in this work can be extended to 
multi-QD case. Since experimentally the mechanism of both the Pauli-blocking and the bichromatic radiation 
approach has been tested and simple manipulation of spin qubits in semiconductor QDs has been  
demonstrated, we believe our scheme to be feasible in the near future.

The authors acknowledge thankfully Ehoud Pazy for his useful comments. MF thanks Jason Twamley for the stimulating discussion about 
the readout problem of solid state quantum computers.

\newpage

\begin{figure}[p]
\begin{center}
\setlength{\unitlength}{1cm}
\begin{picture}(6,8)
\put(-3.5,0){\includegraphics[width=10cm]{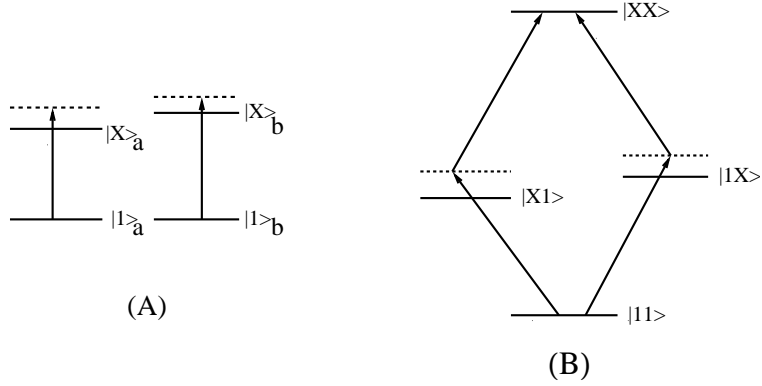}}
\end{picture}
\end{center}
\caption{Two neighboring QDs $a$ and $b$, radiated by two $\sigma^{-}$ polarized lasers, where the arrows represent the laser radiation. 
(A) The Pauli blocking is reflected by the absence of transition between $|0\rangle$ and $|X\rangle$. 
(B) The two-photon process for transition between $|11\rangle$ and $|XX\rangle$ is composed of two blue detunings with respect to QDs $a$
and $b$, respectively, where $|1X\rangle$ and $|X1\rangle$ are non-populated intermediate states.}
\label{Fig1}
\end{figure}

\begin{figure}[p]
\begin{center}
\setlength{\unitlength}{1cm}
\begin{picture}(6,10)
\put(-3.5,0){\includegraphics[width=11cm]{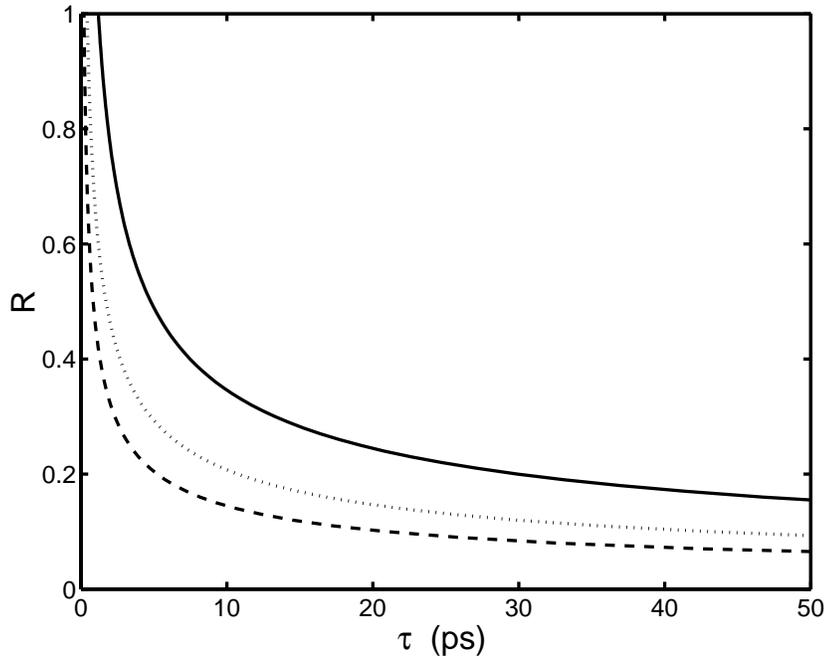}}
\end{picture}
\end{center}
\caption{ The relation between $R=\Omega_{0}/2\delta_{min}$ and $\tau$ in the implementation of a conditional 
phase gate, where the solid, dashed and dotted curves represent the cases of $\delta_{a}=1.5, 2.5,$ and 3.0
$meV$ respectively.}
\label{Fig2}
\end{figure}


\begin{thebibliography}{24}
\bibitem{exp1} The reason we choose the light holes $|\pm 1/2\rangle_{h}$ here is due to the possibility
with $|\pm 1/2\rangle_{h}$ to perform single qubit gate. See \cite{pazy,feng1} for details.
\bibitem{exp2} Numerical results also show that the shortest gating time in our model happens at $\delta=0$, i.e., the case of 
identical QDs. In that case,  the shortest gating time, with $\delta_{a}=2$ meV, is only 6 ps for $R=1/7$. But here we only consider 
different QDs and thereby take $\delta=1$ meV as an example.
\end{thebibliography}
\end{document}